\def\ptitle{\tiny Friedrichs Extensions of Schr\"odinger operators $\dots$}
\font\tr=cmr12                          
\font\it=cmti12                         
\font\trbig=cmbx12 scaled 1500          

\font\tiny=cmr10                        
\output={\shipout\vbox{\makeheadline
                                      \ifnum\the\pageno>1 {\hrule}  \fi 
                                      {\pagebody}   
                                      \makefootline}
                   \advancepageno}

\headline{\noindent {\ifnum\the\pageno>1 
                                   {\tiny \ptitle\hfil
page~\the\pageno}\fi}}
\footline{}

\tr 
\def\bra{{\rm <}}    
\def\ket{{\rm >}}    
\def\nl{\hfil\break\noindent}  
\def\np{\hfil\vfil\break}
\def\hi#1#2{$#1$\kern -2pt-#2} 
\def\hy#1#2{#1-\kern -2pt$#2$} 
\def\htab#1#2{{\hskip #1 in #2}}
\def\dbox#1{\hbox{\vrule 
\vbox{\hrule \vskip #1\hbox{\hskip #1\vbox{\hsize=#1}\hskip #1}\vskip #1 
\hrule}\vrule}} 
\def\qed{\hfill \dbox{0.05true in}} 
\baselineskip 15 true pt  
\parskip=0pt plus 5pt 
\parindent 0.25in
\hsize 6.0 true in 
\hoffset 0.25 true in 
\emergencystretch=0.6 in                 
\vfuzz 0.4 in                            
\hfuzz  0.4 in                           
\vglue 0.1true in
\mathsurround=2pt                        
\topskip=10pt                            
\newcount\zz  \zz=0  
\newcount\q   
\newcount\qq    \qq=0  

\def\pref#1#2#3#4#5{\frenchspacing \global \advance \q by 1     
    \edef#1{\the\q}{\ifnum \zz=1{\item{[{\the\q}]}{#2}{#3}{ #4}{~#5}\medskip} \fi}}

\def\bref #1#2#3#4#5{\frenchspacing \global \advance \q by 1     
    \edef#1{\the\q}
    {\ifnum \zz=1 { %
       \item{[{\the\q}]} 
       {#2}, {#3} {(#4).}{~#5}\medskip} \fi}}

\def\gref #1#2{\frenchspacing \global \advance \q by 1  
    \edef#1{\the\q}
    {\ifnum \zz=1 { %
       \item{[\the\q]} 
       {#2.}\medskip} \fi}}

 \def\sref#1{~[#1]}

\def\references#1{\zz=#1
   \parskip=0pt plus 1pt   
   {\ifnum \zz=1 {\noindent \bf References \medskip} \fi} \q=\qq

\pref{\hala}{R. Hall, N. Saad and A. von Keviczky, Spiked harmonic oscillators. J. Math. Phys. }{43}{(2002) 94-112.}{}

\pref{\halb}{R. Hall, N. Saad and A. von Keviczky, Matrix elements for a generalized spiked harmonic oscilltor. J. Math. Phys. }{39}{(1998) 6345-6351.}{}

\pref{\klau}{J. R. Klauder, Field structure through model studies: Aspects of nonrenormalizable theories, Acta Phys. Austriaca Suppl. }{11}{(1973) 341-387.}{}

\pref{\sim}{B. Simon, Quadratic forms and Klauder's phenomenon: A remark on very singular perturbations, J. Functional Analysis}{14}{(1973) 295-298.}{} 

\pref{\deh}{B. DeFacio and C. L. Hammer, Remarks on the Klauder phenomenon, J. Math. Phys. }{15}{ (1974) 1071-1077.}{}

\pref{\detw}{L. C. Detwiler and J. R. Klauder, Supersingular quantum perturbations,  Phys. Rev. D}{11}{(1975) 1436-1441.}{}

\pref{\eks}{H. Ezawa, J. R. Klauder, and L. A. Shepp, Vestigial effects of singular potentials in duffusion theory and quantum mechanics, J. Math. Phys.}{16}{(1975) 783-799.}{}

\pref{\harr}{E. M. Harrell, Singular perturbation potentials, Ann. Phys. }{105}{(1977)
379-406}{}

\pref{\agua}{V. C. Aguilera-Navarro, G.A. Est\'evez, and R.
Guardiola, Variational and perturbative schemes for a spiked harmonic oscillator,  J. Math. Phys.}{31}{(1990) 99-\#}{}

\bref{\jowa}{J. Weidmann}{Lineare Operatoren in Hilbertr\"aumen}{B. G. Teubner, Stuttgart 1976} {}

\bref{\ruda}{W. Rudin}{Real and Complex Analysis}{McGraw-Hill, New York, 1987}{}

\bref{\doet}{G. Doetsch}{Handbuch der Laplace-Transformation Band I.}{Birkh\"auser Verlag, Basel, 1971}{}

\bref{\rina}{F. Riesz and B. Sz.-Nagy}{Functional Analysis}{F. Ungar Publishing Co., New York, 1955}{p. 335, Thm.}

}

\references{0}    

\htab{3.5}{CUQM-102}

\htab{3.5}{math-ph/0312027}

\htab{3.5}{December 2003}
\vskip 0.5 true in
\centerline{\bf\trbig Friedrichs Extensions of Schr\"odinger Operators}
\vskip 0.1 true in
\centerline{\bf\trbig with Singular Potentials}
\medskip
\vskip 0.25 true in
\centerline{Attila B. von Keviczky$^\dagger$, Nasser Saad$^\ddagger$ and Richard L. Hall$^\dagger$}
\bigskip
{\leftskip=0pt plus 1fil
\rightskip=0pt plus 1fil\parfillskip=0pt\baselineskip 18 true pt
\obeylines
\baselineskip 10 pt
$^\dagger$Department of Mathematics and Statistics, Concordia University,
1455 de Maisonneuve Boulevard West, Montr\'eal, 
Qu\'ebec, Canada H3G 1M8.\par}

\medskip
{\leftskip=0pt plus 1fil
\rightskip=0pt plus 1fil\parfillskip=0pt\baselineskip 18 true pt
\obeylines
\baselineskip 10 pt
$^\ddagger$Department of Mathematics and Statistics,
University of Prince Edward Island, 
550 University Avenue, Charlottetown, 
PEI, Canada C1A 4P3.\par}

\vskip 0.5 true in
\baselineskip 14 pt
\centerline{\bf Abstract}\medskip

The Friedrichs extension for the generalized spiked harmonic oscillator given by the singular differential operator
$
-d^2/dx^2+ Bx^2 + Ax^{-2} + \lambda x^{-\alpha}\ ( B>0, A \geq 0)
$
in $L_2(0,\infty)$ is studied. We look at two different domains of definition for each of these differential operators in $L_2(0,\infty)$, namely $C_0^\infty(0,\infty)$ and 
$D(T_{2,F})\cap D(M_{\lambda, \alpha})$, where the latter is a subspace of the Sobolev space $W_{2,2}(0,\infty)$. Adjoints of these differential operators on $C_0^\infty(0,\infty)$ exist as result of the
null-space properties of functionals. For the other domain, convolutions and Jensen and Minkowski integral inequalities, density of $C_0^\infty(0,\infty)$ in $D(T_{2,F})\cap D(M_{\lambda, \alpha})$ in 
$L_2(0,\infty)$ lead to the other adjoints. Further density properties $C_0^\infty(0,\infty)$ in $D(T_{2,F})\cap D(M_{\lambda, \alpha})$ yield the Friedrichs extension of these differential operators with 
domains of definition $D(T_{2,F})\cap D(M_{\lambda, \alpha})$.
\bigskip
\nl{\bf Keywords}~~Generalized spiked harmonic oscillators, singular potentials, Friedrichs extension, self-adjoint extension, Jensen inequality, Minkowski inequality.
\bigskip

\vfil\eject

\noindent {\bf 1. Introduction}
\medskip
We derive the Friedrichs extension of the singular Hamiltonians (differential operator of the generalized spiked harmonic oscillator \sref{\hala-\halb})
$$
H_\lambda=-d^2/dx^2+ Bx^2 + Ax^{-2} + \lambda x^{-\alpha}\ ( B>0, A \geq 0)\eqno(1)
$$
in the Hilbert space $L_2(0,\infty)$. $H_\lambda$ (1) has been studied as a generalization of the spiked harmonic oscillator \sref{\klau-\agua}
$$
H_0=-{d^2/dx^2}+ Bx^2+l(l+1)x^{-2}+ \lambda x^{-\alpha} \hbox{ ($l$ is the angular momentum number)}.\eqno(2)
$$ 
The generalization (1) lies in $A$ ranging over $[0,\infty)$ instead of $l(l+1),$ $l \in \cal N$. Although the Friedrichs extension of the conventional spiked harmonic oscillator $(A=0)$ has many 
times \sref{\sim-\deh}  been said to exist (the theory of semi-bounded operators \sref{\jowa, Sec. 5.4}), we go much further in the present article; we construct the generalized spiked harmonic 
oscillator's Friedrichs extension for all $A \in [0,\infty)$. 

The Friedrichs extension derives its significance from the following. To provide the spectral decomposition of a symmetric operator $H$ defined {\it in} a Hilbert space ${\cal H},$  whose domain of 
definition $D(H)$ lies dense in ${\cal H},$ namely 
$H = \int\limits_{-\infty}^{\infty}\mu d_{\mu}P_{\mu},$
where $P_{\mu}$ is an ascending \hi{\mu}{parameter} family of projection operators on ${\cal H}$ satisfying $P_{\mu}\rightarrow 0$ or $I$ in the strong sense, according as $\mu\rightarrow -\infty\ \ {\rm or}\ +\infty,$ the operator $H$ must be self adjoint.  Furthermore the projection operators $P_{\mu}$ must satisfy  $P_{\mu+0} = P_{\mu}$ in the strong sense ($P_{\lambda}\rightarrow P_{\mu}$ as $\lambda\downarrow\mu$)\sref{\jowa, p 181, Theorem 7.17},\sref{\rina, Theorem on p. 320}, and the operator $H$ arises out of the inner-product relation
$\bra\  Hf\ |\ g\ \ket = \int_{-\infty}^{\infty}\mu d_{\mu}\bra\  P_{\mu}f\ |\ g\ \ket$ for all $f \in D(H)\quad {\rm and}\quad g \in {\cal H}.$
Usually only a symmetric operator is given, as is the case of our Schr\"odinger operator $H_{\lambda}.$  A self-adjoint extension of $H$ must be found in order that the spectral decomposition of $H$ be applicable.  For the case of our semi-bounded Schr\"odinger operator $H_{\lambda},$ the Friedrichs extension provides this self-adjoint extension on account of the semi-boundedness of $H_{\lambda}.$

There are two principal approaches to the investigation of the domain problem of the Hamiltonian (2). The first is to regard the singular term $x^{-\alpha}$ as a perturbation of the well known harmonic oscillator Hamiltonian, 
the second is to look upon the entire potential as a perturbation of the second order differential operator. To our knowledge, the second approach was never discussed in the literature, while the first 
approach has been investigated by B. Simon \sref{\sim} and DeFacio et al \sref{\deh} and was extensively utilized by Harrell \sref{\harr}.  
The investigation of the generalized spiked harmonic oscillator's Friedrichs extension in the Hilbert space $L_2(0,\infty)$ is accomplished by giving a suitable domain of definition as well as the action of this operator upon elements of this domain of definition. Because we want to maintain the symmetry of this operator in $L_2(0,\infty)$, we must choose as domain a dense linear subspace of $L_2(0,\infty)$, and also, at the same time, guarantee a relatively easy transfer of the operator from the left side of the inner product to the right. As it turns out, it is this transfer of the operator within the inner product that causes difficulty, and shall require the elaboration and derivation of some properties of infinitely differentiable function on $(0,\infty)$ {\it vis-a-vis} derivatives of $L_2(0,\infty)$-functions. 
\medskip\np
\noindent {\bf 2. Analysis of the Hamiltonian $H_\lambda$}
\medskip
Because of the presence of the second derivative operator in the generalized spiked harmonic oscillator (1.1), we note that the second derivative, as an operator in the Hilbert space $L_2(0,\infty)$, 
induces the minimal and maximal second derivative operators $T_{2,0}$ and $T_2$ with domains of definition $D(T_{2,0}) = C_0^\infty(0,\infty)$ and $D(T_2) = W_{2,2}(0,\infty)$ respectively. 
$C_0^\infty(0,\infty)$ is the set of infinitely differentiable complex valued functions on $(0,\infty)$ with compact support, whereas $W_{2,2}(0,\infty)$ designates the Sobolev space consisting of all 
functions $f \in L_2(0,\infty)$ with $f' \in A(0,\infty)$ and $f'' \in L_2(0,\infty)$, where $A(0,\infty)$ is the space of absolutely continuous complex valued functions on $(0,\infty)$. The respective 
actions of these two second derivative operators in the Hilbert space $L_2(0,\infty)$ are: $T_{2,0}f = \tau_2 f \equiv -f''$ and $T_2 f = \tau_2 f$ according as $f \in D(T_{2,0})$ or $f \in D(T_2)$, where 
$\tau_2 = -d^2/dx^2$ acts in the sense of $-(f')'$ almost everywhere on $(0,\infty)$. We moreover have the intermediate situation of the Friedrichs extension $T_{2,F}$ of $T_{2,0}$ with domain of definition and action 
$$
D(T_{2,F}) = \{ f \in W_{2,2}(0,\infty): f(0)=f'(0)=0\},\quad T_{2,F}f = -f''\quad \forall~f \in D(T_{2,F}),\eqno(1)
$$
which is a self-adjoint extension \sref{\jowa, p 157, Theorem 6.32} of $T_{2,0}$ in $L_2(0,\infty)$.

By means of the non-negative parameter $\lambda$ and the second derivative operators $T_{2,0}$ and $T_{2,F}$, the perturbed Hamiltonian operator 
$$
-f''(x) + [B x^2+A x^{-2}+\lambda x^{-\alpha}]f(x),\hbox{ where }f \in D\ (D\hbox{ to be specified}),
$$ 
may be looked upon as an operator in the Hilbert space $L_2(0,\infty)$ in two different ways. First as the operator sum ${\bf H}_\lambda$ of the two operators $T_{2,0}$ and $M_{\lambda;\alpha}$, and second 
also as an operator sum $H_\lambda$, but of $T_{2,F}$ and $M_{\lambda;\alpha}$. At this point we must stress that $M_{\lambda;\alpha}$ stands for the maximal multiplication operator determined 
by the positive continuous (measurable) function $M_{\lambda;\alpha}(x)\equiv B x^2+A x^{-2}+\lambda x^{-\alpha}$ with domain of definition and action given by    
$$
D(M_{\lambda;\alpha})\equiv \{f \in L_2(0,\infty): M_{\lambda;\alpha}f \in L_2(0,\infty)\}~\hbox{ and }~(M_{\lambda;\alpha}f) \equiv M_{\lambda;\alpha}f\quad \forall~ f \in D(M_{\lambda;\alpha})\eqno(2)
$$ 
respectively, where $M_{\lambda;\alpha}f$ is the conventional product of the functions $M_{\lambda;\alpha}$ and $f$. Thus the operators ${\bf H}_\lambda$ and $H_\lambda$ have domains of definition and respective actions given by 
$$
D({\bf H}_\lambda)\equiv D(T_{2,0})\cap D(M_{\lambda;\alpha}) = C_0^\infty(0,\infty),\quad
{\bf H}_\lambda f \equiv -f'' + M_{\lambda;\alpha}f \quad \forall~f \in D({\bf H}_\lambda);\eqno(3a)
$$
$$\eqalign{&
D(H_\lambda)\equiv D(T_{2,F})\cap D(M_{\lambda;\alpha})= \{f \in L_2(0,\infty):f' \in A(0,\infty);f'', M_{\lambda;\alpha}f
\in L_2(0,\infty)\},\cr
&H_\lambda f \equiv -f'' + M_{\lambda;\alpha}f \quad \forall~f \in D(H_\lambda).}\eqno(3b)
$$
By denoting the inner product of two elements $f$ and $g$ in the Hilbert space $L_2(0,\infty)$ by $\bra f\ |\ g\ket \equiv \int_0^\infty f(x)\overline{g(x)}dx$, we have, that ${\bf H}_\lambda$ and $H_\lambda$ are symmetric as well as semi-bounded from below, since both $D({\bf H}_\lambda)$ and $D(H_\lambda)$ lie dense in $L_2(0,\infty)$. These facts are direct consequences of:  
$D(H_\lambda)\supset C_0^\infty(0,\infty)$, $C_0^\infty(0,\infty)$ is dense in $L_2(0,\infty)$,
$$
\bra H_\lambda f\ |\ g \ket  = \bra  -f''\ |\ g \ket  + \bra  M_{\lambda;\alpha}f\ |\ g \ket  = \bra  f'\ |\ g' \ket  + \bra  M_{\lambda;\alpha}f\ |\ g \ket  = \bra  f\ |\ H_\lambda g \ket\eqno(4)  
$$
and
$$
\bra {\bf H}_\lambda f\ |\ g \ket  = \bra  -f''\ |\ g \ket  + \bra  M_{\lambda;\alpha}f\ |\ g \ket  = \bra  f'\ |\ g' \ket  + \bra  M_{\lambda;\alpha}f\ |\ g \ket  = \bra  f\ |\ {\bf H}_\lambda g \ket\eqno(5)  
$$ 
for all $f, g \in D(H_\lambda)$ and $D({\bf H}_\lambda)$ respectively. In equations (4) and (5), the last two equalities  are an immediate result of: integration by parts; $C_0^\infty(0,\infty)$-functions 
always have compact subsets of $(0,\infty)$ for supports, as well as the property that  $f(0^+)$ and $f'(0^+)$ always exist \sref{\jowa, p 153, Theorem 6.27} with $f(\infty) = f'(\infty) = 0$ for any 
$f \in W_{2,2}(0,\infty)$. We further point out that the maximal multiplication operator $M_{\lambda;\alpha}$ is the linear combination of the maximal multiplication operators $M$ and $(\cdot)^\alpha$ determined by the 
continuous (measurable) functions $M(x) = B x^2+A x^{-2}$ and $(\cdot)^\alpha(x) = x^{-\alpha}$ on $(0,\infty)$ respectively. Their corresponding domains of definition are 
$$
D(M) = \{f \in L_2(0,\infty): Mf \in L_2(0,\infty)\}
\hbox{ and }
D((\cdot)^\alpha) = \{f \in L_2(0,\infty): (\cdot)^\alpha f \in L_2(0,\infty)\}\eqno(6) 
$$
with self-evident actions in terms of the conventional multiplication of functions. If we define the unperturbed Hamiltonians ${\bf H}_0 \equiv T_{2,0} + M$ and $H_0 \equiv T_{2,F} + M$ in the sense of operator addition in the Hilbert space $L_2(0,\infty)$, then the perturbed Hamiltionians assume the forms 
$$
{\bf H}_\lambda = {\bf H}_0 + \lambda (\cdot)^\alpha\hbox{ and }H_\lambda = H_0 + \lambda (\cdot)^\alpha\eqno(7)
$$ 
of the sum two operators in $L_2(0,\infty)$ in terms of the perturbation parameter $\lambda$.

\medskip
\noindent {\bf 3. The Adjoint of ${\bf H}_\lambda$}\medskip

We now consider the adjoints of the $\lambda$-parameter family of Hamiltonians 
${\bf H}_\lambda$ in the Hilbert space $L_2(0,\infty)$, whose domains of definition and accompanying actions are 
$$
D({{\bf H}_\lambda}^\dagger) \equiv \{g \in L_2(0,\infty): \exists\ {\hat g} \ni \bra  {\bf H}_\lambda f\ |\ g \ket  = \bra  f\ |\ \hat{g} \ket  \forall f \in D({\bf H}_\lambda)\}\hbox{ and }{{\bf H}_\lambda}^\dagger g = \hat{g},\eqno(1)
$$
where the linear manifold $D({\bf H}_\lambda)$ is replaceable by $C_0^\infty(0,\infty)$ in consequence of $D({\bf H}_\lambda) = C_0^\infty(0,\infty)$. For $g \in D({{\bf H}_{\lambda}}^\dagger)$ and $f \in C_0^\infty(0,\infty)$ we have that 
$$
\bra  {\bf H}_\lambda f\ |\ g \ket  = \bra  -f''\ |\ g \ket  + \bra  M_{\lambda;\alpha}f\ |\ g \ket  = \bra  f\ |\ {{\bf H}_\lambda}^\dagger g \ket ,\eqno(2)
$$
and because ${{\bf H}_\lambda}^\dagger g - M_{\lambda;\alpha} \in L_2^{loc}(0,\infty)$, we may rewrite this equation as 
$$
\bra  -f''\ |\ g \ket  = \bra  f\ |\ {{\bf H}_\lambda}^\dagger g - M_{\lambda;\alpha}g \ket  = \int_0^\infty f(x)\overline {[({{\bf H}_\lambda}^\dagger g)(x) - (M_{\lambda;\alpha}g)(x)]}dx.\eqno(3)
$$
Let $c$ be a fixed positive number, then the function 
$$
h_\lambda (x) = -\int_c^x \int_c^{x_1}({{\bf H}_\lambda}^\dagger g - M_{\lambda;\alpha}g)(x_2)dx_2dx_1 = -\int_c^x  (x - x_2)({{\bf H}_\lambda}^\dagger g - M_{\lambda;\alpha}g)(x_2)dx_2\eqno(4)
$$ 
is continuously differentiable with absolutely continuous ${h_\lambda}'$ on $(0,\infty)$ - i. e.  ${h_\lambda}' \in A(0,\infty)$ - and 
$$
(\tau_2 h_\lambda)(x) = ({{\bf H}_\lambda}^\dagger g - M_{\lambda;\alpha}g)(x)\hbox{ for almost all }x \in (0,\infty).\eqno(5) 
$$
Moreover, we point to the fact that  
$$
\bra  f\ |\ {{\bf H}_\lambda}^\dagger g - M_{\lambda;\alpha}g \ket  = \bra  f\ |\ \tau_2 h_\lambda \ket  = \bra  f'\ |\ h_\lambda' \ket  = \bra  - f''\ |\ h_\lambda \ket \ \forall f\in C_0^\infty(0,\infty),\eqno(6)
$$ 
whence $\bra  -f''\ |\ g \ket  = \bra -f''\ |\ h_\lambda \ket $ or $\bra -f''\ |\ g - h_\lambda \ket  = 0$ for all $f \in C_0^\infty(0,\infty)$. Therefore,  $\bra  \cdot\ |\ g - h_\lambda \ket $ determines a linear functional 
$$
L: C_0^\infty(0,\infty) \mapsto {\cal C}\hbox{ with }L(f) = \bra  f\ |\ g - h_\lambda \ket  = \int_0^\infty f(x)\overline{[g(x) - h_\lambda (x)]}dx,\eqno(7) 
$$ 
whose null space satisfies, in terms of the elementary linear functionals $L_j: C_0^\infty(0,\infty) \mapsto {\cal C}$ with $L_j(f)$ defined \sref{\jowa, p 154, Theorem 6.28} as the integral of the function 
$x^jf(x)$ over the interval $(0,\infty)$ ($j=0,1$),
$$
N(L)\supset Range(T_{2,0}) = \{f \in C_0^\infty(0,\infty): L_j(f) = 0\hbox{ for } j=0,1\} = N(L_0)\cap N(L_1).\eqno(8)
$$
However, $N(L_0)\cap N(L_1) \subset N(L)$ entails that \sref{\jowa, p 56, Theorem 4.1} 
$$
L = \overline{c_0}L_0 + \overline{c_1}L_1 \hbox{ or }L(f) = \bra  f\ |\ g - h_\lambda \ket  = \int_0^\infty f(x)\overline {[c_0 + c_1x]}dx\ \forall f \in C_0^\infty(0,\infty),\eqno(9)
$$
namely $(g - h_\lambda)(x) = c_0 + c_1x$ a.e. on $(0,\infty)$ or in terms of the previously defined $h_\lambda$ 
$$
g(x) = -\int_c^x \int_c^{x_1}({{\bf H}_\lambda}^\dagger g - M_{\lambda;\alpha}g)(x_2)dx_2dx_1 + c_0 + c_1x\ \forall x \in(0,\infty).\eqno(10)
$$
Thus $g'$ is absolutely continuous on $(0,\infty)$, $-g'' = h_\lambda'' = {{\bf H}_\lambda}^\dagger g - M_{\lambda;\alpha}g$ as well as $g'' \in L_2^{loc}(0,\infty)$; in particular $-g'' + M_{\lambda;\alpha}g = {{\bf H}_\lambda}^\dagger g$. We have therefore 
\medskip
\noindent{\bf LEMMA 3.1:}
The adjoint ${{\bf H}_\lambda}^\dagger$ of ${{\bf H}_\lambda}$ in the Hilbert space $L_2(0,\infty)$ has domain of definition and action given by 
$$
D({{\bf H}_\lambda}^\dagger) = \bigg\{g \in L_2(0,\infty): g' \in A(0,\infty),\int_0^\infty|(-g'' + M_{\lambda;\alpha}g)(x)|^2dx\bra \infty \bigg\}
$$
$$
\hbox{and }{{\bf H}_\lambda}^\dagger g = (-g'' + M_{\lambda;\alpha}g).\eqno(11)
$$ 
\medskip
\noindent{PROOF.}
$M_{\lambda,\alpha}(x)$ continuous on $(0,\infty)$ and $(-g'' + M_{\lambda;\alpha}g) \in L_2(0,\infty)$ implies that $g'' \in L_2^{loc}(0,\infty)$; therefore, $g'' \in L_2^{loc}(0,\infty)$ is 
omitted from within curly bracket. On the other hand, if $g$ belongs to the set within the curly brackets, then $g$ belongs to the domain $D({{\bf H}_\lambda}^\dagger)$ of ${{\bf H}_\lambda}^\dagger$. 
This is evident from taking inner products with $C_0^\infty(0,\infty)$-functions.\qed 
\medskip

We note that $(-g'' + M_{\lambda;\alpha}g)$ must be taken collectively, and not as the sum  $-g'' + M_{\lambda;\alpha}g$ of two $L_2(0,\infty)$-functions arising out of the operator sum of $\tau_2$ 
and $M_{\lambda;\alpha}$. The adjoint ${{\bf H}_\lambda}^\dagger \supset {\bf H}_\lambda$, but ${{\bf H}_\lambda}^\dagger \neq {\bf H}_\lambda$, because  for some $g \in D({{\bf H}_\lambda}^\dagger)$, 
$g''$ and $M_{\lambda;\alpha}g$ fail to belong to $L_2(0,\infty)$, but their difference $(-g'' + M_{\lambda;\alpha}g)\in L_2(0,\infty)$, as the following counter-example 
indicates. We define the function $\psi (x) \equiv x^{1/2 - \sqrt {1/4 + A}} \phi(x)$ ($0 < A < 3/2$), where $\phi$ is the restriction of 
a $C_0^\infty ({\cal R})$-function $\Phi$ to $(0,\infty)$, such that $\Phi(x)=1$ in a neighborhood of the origin. 
$$
(M_{\lambda;\alpha}\psi)(x) = [Bx^{5/2 - \sqrt {1/4 + A}} + Ax^{-3/2 - \sqrt {1/4 + A}} + \lambda x^{1/2 - \sqrt {1/4 + A}- \alpha}]
\phi(x)\quad(0 < \alpha < \sqrt {1/4 + A}) 
$$
fails to be an $L_2(0,\infty)$-function; further, 
$$
\psi''(x) = -A x^{-3/2 - \sqrt {1/4 + A}}\phi(x) + 
(1 - \sqrt {1 +4 A})x^{-1/2 - \sqrt {1/4 + A}}\phi'(x) + x^{1/2 - \sqrt {1/4 + A}}\phi''(x),\eqno(12)
$$
where  $\phi'(x) = \phi''(x) = 0$ in a neighborhood about the origin. $\psi'' \notin L_2(0,\infty)$; however, 
$$\eqalign{
({{\bf H}_\lambda}^\dagger \psi)(x) &= -\psi''(x) + (M_{\lambda;\alpha} \psi)(x) = (\sqrt{1 + 4A} - 1)x^{-1/2 - \sqrt {1/4 + A}}\phi'(x)\cr 
& + x^{1/2 - \sqrt {1/4 + A}}\phi''(x) +[Bx^{5/2 - \sqrt {1/4 + A}}+ \lambda x^{1/2 - \sqrt {1/4 + A}- \alpha}]\phi(x)}\eqno(13)
$$ 
belongs to $L_2(0,\infty)$ (the singularities at $0$ cancel as result of the presence of $\phi'$ and $\phi''$. 
\medskip
\noindent {\bf 4. Special Density Properties of $C_0^\infty(0,\infty)$}
\medskip
Since the operator ${\bf H}_\lambda$ in $L_2(0,\infty)$ with $D({\bf H}_\lambda) = C_0^\infty(0,\infty)$ and 
${\bf H}_\lambda f = T_{2,0}f + M_{\lambda,\alpha}f$ has the Hilbert space adjoint ${{\bf H}_\lambda}^\dagger$ as defined  in Lemma 3.1, 
it is also possible to ascertain the adjoint ${H_\lambda}^\dagger$ of the operator 
$H_\lambda$ with $D(H_\lambda) \equiv D(T_{2,F})\cap D(M_{\lambda,\alpha})$ and $H_\lambda f \equiv (T_{2,F} + M_{\lambda,\alpha})f$ in 
terms of the sum of the two operators $T_{2,F}$ and $M_{\lambda,\alpha}$ in $L_2(0,\infty)$. Further, the transfer of $H_\lambda^\dagger$ 
from the right side of $< f\ |\ H_\lambda^\dagger g >$ to the left in $< H_\lambda f\ |\ g >$ for all $f \in D(H_\lambda)$ and 
$g \in D({H_\lambda}^\dagger)$ requires integration by parts, which worked so nicely for $< {\bf H_\lambda} f\ |\ g > = 
< f\ |\ {{\bf H_\lambda}}^\dagger g >,$ causes insurmountable difficulties in the case of $f \in D(H_\lambda)$ and 
$g \in D({{H_\lambda}}^\dagger)$. This is because $g \in D({H_\lambda}^\dagger)$ guarantees merely  $g'' \in L_2^{loc}(0,\infty)$, and 
thus an {\it a priori} statement concerning $g(0)$ and $g'(0)$ is not possible. Therefore, we shall proceeded in another way.

We note that every $f \in D(T_{2,F})$ is continuous on $[0,\infty)$ and has  absolutely continuous $f'$ on $[0,\infty)$ with 
$f(0) = f'(0) = 0$; hence, a shift of function values by $a>0$ to the right, yields a $D(T_{2,F})$-function again. Thus the function $F_a$ 
$$
F_a(x) \equiv 0\hbox{ for }0 \leq x < a\hbox{ and }\equiv f(x-a)\hbox{ for }x \geq a\ \forall f\in D(T_{2,F}).\eqno(1) 
$$
obtained from $f$ belongs to $D(T_{2,F})$. Out of $F_a \in L_2(0,\infty)$, ${F_a}'(x) = 0$ for $0 \leq x < a$ and $ = f'(x-a)$ for 
$x \geq a$, ${F_a}''(x) = 0$ for $0 < a$ and $ = f''(x-a)$ for $x > a$ readily lead to the properties: $F_a(0) = {F_a}'(0) =0$, 
$F_a \in L_2(0,\infty)$, and ${F_a}'$ is absolutely continuous with ${F_a}'' \in L_2(0,\infty)$. This triplet of functions $\{F_a, {F_a}', {F_a}'' \}$ satisfies \sref{\ruda, p 182, Theorem 9.5}: 
$$ 
\|f^{(j)} - {F_a}^{(j)}\| \rightarrow 0\hbox{ as }a \rightarrow 0^+\ (j = 0, 1, 2),\eqno(2) 
$$
where these three functions are extendable to all of ${\cal R}$ by $F_a(x) \equiv 0$ ($x < 0$). Each of this triplet we ``smoothen'' by 
convoluting them with the $C_0^\infty({\cal R})$-function    
$$
\delta_\eta(x) \equiv A_\eta exp ((x^2 - \eta^2)^{-1})\hbox{ for }|x|< \eta\hbox{ and }\equiv 0\hbox{ for }|x|\geq a,\hbox{ where } 
\int_{-\infty}^\infty \delta_\eta(s)ds=1\eqno(3)
$$
for values of $\eta < a/8$. At this stage we point out that $\delta_\eta(x)dx$ defines a positive measure on $(-\eta,\eta)$ whose measure 
of the interval $(-\eta,\eta)$ is $1$; this permits the utilization of the Jensen integral inequality \sref{\ruda, p 62, Theorem 3.3}. 
In consequence of $\delta_\eta(\pm \eta) = {\delta_\eta}'(\pm \eta) = 0$, these ``smoothened'' functions satisfy 
\sref{\doet, p 120, theorem 14}: 
$$(F_a \ast \delta_\eta)' = F_a \ast {\delta_\eta}' = {F_a}' \ast \delta_\eta\hbox{ and }(F_a \ast \delta_\eta)'' = 
F_a \ast {\delta_\eta}'' = {F_a}'' \ast \delta_\eta.$$
Turning to the first three derivatives of $F_a$, we have from 
$$
|{F_a}^{(j)}(x) - ({F_a}^{(j)} \ast \delta_\eta)(x)| \leq \int_{-\eta}^\eta |{F_a}^{(j)}(x) - {F_a}^{(j)}(x - t)|\delta_\eta(t)dt
\ (\eta < a/8; j=0,1,2)\eqno(4)
$$ 
and Jensen's integral inequality ($u^2$ being convex on ${\cal R}$) that 
$$
|{F_a}^{(j)}(x) - ({F_a}^{(j)}\ast \delta_\eta)(x)|^2 \leq \int_{-\eta}^\eta |{F_a}^{(j)}(x) - {F_a}^{(j)}(x - t)|^2\delta_\eta(t)dt
\ (j=0,1,2).\eqno(5)
$$
The Minkowski integral inequality \sref{\ruda, p 182, Prob. 19} applied to inequality (5) yields:
$$
\|{F_a}^{(j)} - ({F_a}^{(j)}\ast \delta_\eta)\| \leq \int_{-\eta}^\eta \|{F_a}^{(j)} - {F_{a+t}}^{(j)}\|\delta_\eta(t)dt\rightarrow 0\hbox{ for }\eta\rightarrow 0^+ \ (j = 0, 1 , 2),\eqno(6)
$$
because each of the maps $t\mapsto {F_t}^{(j)}$ constitutes \sref{\ruda, p 182, Theorem 9.5} a continuous map  $(0, \infty) \mapsto 
L_2(0,\infty)$ ($j=0,1,2$). Herefrom the triplet of functions $\{(F_a\ast\delta_\eta) , (F_a\ast\delta_\eta)', 
(F_a\ast\delta_\eta)'' \}$ satisfies 
$$
\|{F_a}^{(j)} - ({F_a}^{(j)}\ast \delta_\eta)\| \rightarrow 0\hbox{ for }\eta\rightarrow 0^+ \ (j =0,1,2).\eqno(7)
$$ 
 
We further proceed by introducing the $C_0^\infty({\cal R})$-function $\chi_{[-R-r,R+r]}\ast \delta_r$ for$R, r > 0$, where 
$\chi_{[-R-r,R+r]}\ast \delta_r$ assumes on $[-R,R]$ and ${\cal R}\setminus [-R-2r,R+2r]$ the values of $1$ and $0$ respectively, and 
maps each of the intervals $[-R-2r,-R)$ and $(R,R+2r]$ monotonely onto $[0,1)$. Let $\varphi_{R,r}$ be the restriction of 
$\chi_{[-R-r,R+r]}\ast \delta_r$ to $[0,\infty)$. Thus $\varphi_{R,r}([0,R])= \{ 1 \}$, $\varphi_{R,r}((R,R+2r])= [0,1)$, 
$\varphi_{R,r}([R+2r,\infty))= \{ 0 \}$, and ${\varphi_{R,r}}'(x) = -\delta_r(x-R-r)$ is a $C_0^\infty(0,\infty)$-function with 
$supp({\varphi_{R,r}}') = [R,R+2r]$. Its effect on the triplet $\{(F_a \ast \delta_\eta),(F_a \ast \delta_\eta)',
(F_a \ast \delta_\eta)''\}$ is as follows. First, $(F_a \ast \delta_\eta)\varphi_{R,r}$ is a $C_0^\infty(0,\infty)$-function, for which
$$
\|F_a \ast \delta_\eta - (F_a \ast \delta_\eta)\varphi_{R,r}\| = \|(F_a \ast \delta_\eta)[1 -\varphi_{R,r}]\| \leq \|(F_a \ast \delta_\eta)\chi_{[R,\infty)}\| \rightarrow 0\hbox{ for }R\rightarrow \infty\eqno(8) 
$$ 
implies  
$$
\|F_a \ast \delta_\eta - (F_a \ast \delta_\eta)\varphi_{R,r}\| \rightarrow 0\hbox{ for }R\rightarrow \infty.\eqno(9)
$$
Second, the chain of inequalities
$$\eqalign{&
\|(F_a \ast \delta_\eta)' - ((F_a \ast \delta_\eta)\varphi_{R,r})'\| = \|({F_a}' \ast \delta_\eta)[1 - \varphi_{R,r}] 
-(F_a \ast \delta_\eta){\varphi_{R,r}}'\|\leq 
\|({F_a}' \ast \delta_\eta)[1 - \varphi_{R,r}]\|\cr &  + \|(F_a \ast \delta_\eta){\varphi_{R,r}}'\| \leq 
\|({F_a}' \ast \delta_\eta)\chi_{[R,\infty)}\| + \| \delta_\eta \|_\infty||(F_a \ast \delta_\eta)\chi_{[R,r+2r]}\| \rightarrow 0
\hbox{ for }R\rightarrow \infty}\eqno(10) 
$$
entails that
$$
\|(F_a \ast \delta_\eta)' - ((F_a \ast \delta_\eta)\varphi_{R,r})'\|\rightarrow 0\hbox{ for }R\rightarrow \infty.\eqno(11)  
$$
And thirdly, by estimating as follows
$$\eqalign{
\|(F_a \ast \delta_\eta)''& - ((F_a \ast \delta_\eta)\varphi_{R,r})''\|= 
\|{F_a}'' \ast \delta_\eta - ({F_a}'' \ast \delta_\eta)\varphi_{R,r} - 2({F_a}' \ast \delta_\eta){\varphi_{R,r}}' - 
(F_a \ast \delta_\eta){\varphi_{R,r}}''\|
\cr
& \leq \|({F_a}'' \ast \delta_\eta)[1 - \varphi_{R,r}]\| + 2\|({F_a}' \ast \delta_\eta){\varphi_{R,r}}'\| + 
\|(F_a \ast \delta_\eta){\varphi_{R,r}}''\|\cr
& \leq
\|({F_a}'' \ast \delta_\eta)\chi_{[R,\infty)}\| + 2\| \delta_\eta \|_\infty \|({F_a}' \ast \delta_\eta)\chi_{[R,r+2r]}\| +
\| {\delta_\eta}'\|_\infty \|(F_a \ast \delta_\eta)\chi_{[R,r+2r]}\|}\eqno(12)                     
$$
we arrive at  
$$
\|(F_a \ast \delta_\eta)'' - ((F_a \ast \delta_\eta)\varphi_{R,r})''\| \rightarrow 0\quad\hbox{ for }\quad R\rightarrow \infty.\eqno(13)
$$
In summary we have that the triplet of functions 
$\{(F_a\ast\delta_\eta)\varphi_{R,r},   ((F_a\ast\delta_\eta)\varphi_{R,r})', ((F_a\ast\delta_\eta)\varphi_{R,r})'' \}$
satisfies
$$
\|(F_a \ast \delta_\eta)^{(j)} - ((F_a \ast \delta_\eta)\varphi_{R,r})^{(j)}\| \rightarrow 0\quad\hbox{ for }\quad R\rightarrow \infty \ (j =0,1,2).\eqno(14)
$$
Let us now turn to the maximal multiplication operator $M_{\lambda,\alpha}$ in $L_2(0,\infty)$ by deriving more general properties of the 
function $F_a$ constructed from $f \in D(H_\lambda)$. Specifically, let $\mu(x)$ be non-negative continuous function on $(0,\infty)$, 
which is strictly decreasing and strictly increasing on $(0,r_\mu]$ and $[R_\mu, \infty)$ respectively with $r_\mu \leq R_\mu$, and in 
addition satisfies  
$$
\lim_{x \rightarrow \infty} {{\mu (x+1)}\over {\mu(x)}} = C_\mu < \infty.\eqno(15)
$$
\medskip
\noindent{\bf LEMMA 4.1:}
$\sqrt{\mu}f \in L_2(0,\infty)$ implies that $\sqrt{\mu}F_a \in L_2(0,\infty)$ for all $a \geq 0$. 
\medskip
\noindent{PROOF.}
The validity of this statement follows from first trapping $a$ between two successive non-negative integers, namely $n \leq a < n+1$, 
and estimating thereafter as follows:
$$\eqalign{
\int_0^\infty |F_a(x)|^2\mu(x)dx &= \int_0^\infty |f(x)|^2\mu(x + a)dx\leq 
\int_0^{R'} |f(x)|^2\mu(x + a)dx + \int_{R'}^\infty |f(x)|^2\mu(x + n + 1)dx\cr
& =
\int_0^{R'} |f(x)|^2\mu(x + a)dx + \int_{R'}^\infty |f(x)|^2\bigg[{\mu(x + n + 1)\over \mu(x)}\bigg]\mu(x)dx\cr
& =   
\int_0^{R'} |f(x)|^2\mu(x + a)dx + \int_{R'}^\infty |f(x)|^2\bigg\{\prod\limits_{k=0}^n{\mu(x + k + 1)\over \mu(x + k)}\bigg\}\mu(x)dx \cr
&\leq
\int_0^{R'} |f(x)|^2\mu(x + a)dx + [1 + C_\mu]^{n+1}\int_{R'}^\infty |f(x)|^2\mu(x)dx < \infty,}\eqno(16)
$$
where $R'>R_\mu$ and is picked so large that ${\mu(x+1)\over \mu(x)} \leq 1 + C_\mu$ for all $x \geq R'$.\qed
\medskip
Estimating $\|\sqrt{\mu}f - \sqrt{\mu}F_a\|$, for sufficiently small $a$, leads to the following
\medskip
\noindent{\bf LEMMA 4.2:}

\noindent For $f \in D(T_{2,F})$ satisfying $\sqrt{\mu}f \in L_2(0,\infty)$, we have that $\|\sqrt{\mu}f - \sqrt{\mu}F_a\|\rightarrow 0$ as $a\rightarrow 0^+$.
\medskip
\noindent{PROOF.}
We make a restriction on $a$, namely $0 < a < \min\{r_\mu, 1\}$, and in terms of a positive $r < r_\mu$, which remains unspecified 
at this moment, we estimate as follows
$$\eqalign{&
\|\sqrt{\mu}f - \sqrt{\mu}F_a\| = \|[\sqrt{\mu}f - \sqrt{\mu}F_a]\chi_{(0,r]}\| + \|[\sqrt{\mu}f - \sqrt{\mu}F_a]\chi_{(r,\infty)}\|\cr
& \leq  
\|(\sqrt{\mu}f)\chi_{(0,r]}\| + \|(\sqrt{\mu}F_a)\chi_{(0,r]}\|+ \|\sqrt{\mu}(f - F_a)\chi_{(r,\infty)}\|\cr
& \leq  
\|(\sqrt{\mu}f)\chi_{(0,r]}\| + \|(\sqrt{\mu}F_a)\chi_{(0,r]}\| + \|\sqrt{\mu}(f - F_a)\chi_{(r,b+1]}\| + \|\sqrt{\mu}(f - F_a)\chi_{(b+1,\infty)}\|\cr
&\leq 
\|(\sqrt{\mu}f)\chi_{(0,r]}\| + \|(\sqrt{\mu}F_a)\chi_{(0,r]}\| + 
\|\sqrt{\mu}(f - F_a)\chi_{(r,b+1]}\| + \|(\sqrt{\mu}f)\chi_{(b+1,\infty)}\|+ \|(F_a)\chi_{(b+1,\infty)}\|}\eqno(17)
$$
We further estimate the last norm expression, by replacing the variable of integration $x$ by $x+1$ under the condition $0 < a < \min\{r_\mu, 1\}$, and lets us arrive at
$$
\eqalign{\|(\sqrt{\mu}F_a)\chi_{[b+1,\infty)}\|^2 &= \int_{b+1-a}^\infty |f(x)|^2\mu(x+a)dx\leq \int_b^\infty |f(x)|^2\mu(x+1)dx\cr
& = 
\int_b^\infty |f(x)|^2[\mu(x+1)/\mu(x)]\mu(x)dx \leq [1 + C_\mu]\int_b^\infty |f(x)|^2\mu(x)dx}\eqno(18)
$$
where $(b \geq R')$ with $R'$ the same as in Lemma 4.1. We thus have for the previous norm estimate in terms of the immediately preceeding integral estimate that
$$\eqalign{&
\|\sqrt{\mu}f - \sqrt{\mu}F_a\| \leq \sqrt{\int_0^r |f(x)|^2\mu(x)dx} + \sqrt{\int_0^r |F_a(x)|^2\mu(x)dx} 
\cr &+ \sqrt{\int_r^{b+1} |f(x) - f(x - a)|^2\mu(x)dx} + \bigg[ 1 + \sqrt{1 + C_\mu}\bigg]\sqrt{\int_b^\infty |f(x)|^2\mu(x)dx}\quad (b \geq R').}\eqno(19)
$$ 
Now we turn to choosing $a$ sufficiently small so that $\|\sqrt{\mu}f - \sqrt{\mu}F_a\| < \epsilon$.  Because $\|\sqrt{\mu}f\| < \infty$, we can deduce the existence of $r(\epsilon)\leq r_\mu$ and $R(\epsilon)$ such that  
$$r < r(\epsilon)\hbox{ and }\rho > R(\epsilon)\Rightarrow  \sqrt{\int_0^r |f(x)|^2\mu(x)dx},\ \bigg[1 + \sqrt{1 + C_\mu}\bigg]\sqrt{\int_\rho^\infty |f(x)|^2\mu(x)dx} < \epsilon/4,\eqno(20)
$$ 
whereby the first and fourth integral expressions in inequality (19) are simultaneously less than $\epsilon/4$ provided $r < r(\epsilon)$ and $b > R(\epsilon)$. 

If $r < r(\epsilon)$, then the second integral expression in inequality (19) is less than $\epsilon/4$, as can be easily seen from the following calculations. This second integral expression is zero if $r \leq a$ and thus 
$$\eqalign{&
\sqrt{\int_0^r |F_a(x)|^2\mu(x)dx} = \chi_{[0,r)}(a)\sqrt{\int_a^r |f(x -a)|^2\mu(x)dx}=  
\chi_{[0,r)}(a)\sqrt{\int_0^{r-a}|f(x)|^2\mu(x+a)dx}\cr
&\leq
\chi_{[0,r)}(a)\sqrt{\int_0^{r-a}|f(x)|^2\mu(x)dx} \leq \chi_{[0,r)}(a)\sqrt{\int_0^r|f(x)|^2\mu(x)dx}\leq \chi_{[0,r)}(a)\epsilon/4,}\eqno(21)
$$
because $r < r(\epsilon)$ implies $r\leq r_\mu$ (choice of $r(\epsilon)$) and $\mu(x)$ decreasing on $(0,r_\mu]$ leads to $\mu_(x + a) \leq \mu(x)$ since $x+a \leq r - a + a = r$.   

By fixing an $r < r(\epsilon)$ and a $b > max\{R',R(\epsilon)\}$, we guarantee that the sum of the first, second, and fourth integral 
expressions in inequality (19) is less than $3\epsilon/4$. In the third integral expression in inequality (19) $f$ is continuous on $[0,\infty)$, hence uniformly continuous on the closed interval $[0,b+1]$. Therefore, there exists a $\Delta(\epsilon)$ such that 
$$
x,y \in [0,b+1]\hbox{ with }|x-y|< \Delta(\epsilon)\Rightarrow|f(x)-f(y)| < \epsilon/4.\eqno(22)$$
Setting $\mu_b \equiv \max\{\mu(x): r \geq x \leq b+1 \}$, which quantity is non-negative, and thereafter choosing $a < \Delta(\epsilon[4^2(b+1)(\mu_b+1)]^{-1})$, leads us to
$$
\sqrt{\int_r^{b+1} |f(x) - f(x - a)|^2\mu(x)dx} < \epsilon/4;\eqno(23)
$$
whereby the proof is complete.\qed
\medskip
\noindent{\bf LEMMA 4.3:}
For $f \in D(T_{2,F})$ satisfying $\sqrt{\mu}f \in L_2(0,\infty)$, we have for each $a >0$ that $\|\sqrt{\mu}F_a - \sqrt{\mu}F_{a+t}\|\rightarrow 0$ as $t\rightarrow 0$.
\medskip
\noindent{PROOF.}
Since Lemma 4.2 maintains that $\|\sqrt{\mu}f - \sqrt{\mu}F_a\|\rightarrow 0$ as $a\rightarrow 0^+$, we are tempted to conclude from Lemma 4.2 the validity of the assertion for $t\rightarrow 0$. Unfortunately, we may only conclude the validity for $t\rightarrow 0^+$ on account of $\sqrt{\mu}F_a \in L_2(0,\infty)$. Consequently, we start afresh. The positive quantity $a$ we hold fixed, and trap it between two successive integers, as in the proof of Lemma 4.1, namely $n \leq a < n+1$, and only admit $t$ satisfying $|t| \leq \min\{1, a/4\}$; thus, $3a/4 \leq a+t < n+2$ and $F_a(x) = F_a(x-t)$ for $0 \leq x \leq a/2$. We write
$$\eqalign{
\|\sqrt{\mu}F_a &- \sqrt{\mu}F_{a+t}\| = \|\sqrt{\mu}(F_a - F_{a+t})\chi_{[a/2,R+2]}\| + \|\sqrt{\mu}(F_a - F_{a+t})\chi_{(R+2,\infty]}\|\cr
& \leq
\|\sqrt{\mu}(F_a - F_{a+t})\chi_{[a/2,R+2]}\| + \|(\sqrt{\mu}F_a)\chi_{(R+2,\infty]}\| + \|(\sqrt{\mu}F_{a+t})\chi_{(R+2,\infty]}\|}\eqno(24) 
$$
and estimate the last two norm expressions as follows:
$$\eqalign{&
\|(\sqrt{\mu}F_a)\chi_{(R+2,\infty]}\| = \bigg[\int_{R+2}^\infty |f(x-a)|^2\mu(x)dx \bigg]^{1/2}=\bigg[\int_{R+2-a}^\infty |f(x)|^2\mu(x+a)dx\bigg]^{1/2}\cr
&\leq
\bigg[\int_{R+2-a}^\infty |f(x)|^2\mu(x+n+2)dx\bigg]^{1/2}=  \bigg[\int_{R+2-a}^\infty |f(x)|^2\bigg\{ \prod_{k=0}^{n+1}{\mu(x+k+1)\over \mu(x+k)}\bigg\}\mu(x)dx\bigg]^{1/2}\cr
& \leq 
(1 + C_\mu)^{1+n/2}
\bigg[\int_{R+2-a}^\infty |f(x)|^2\mu(x)dx\bigg]^{1/2}\leq (1 + C_\mu)^{1+n/2}||(\sqrt{\mu}f)\chi_{(R+1-a,\infty]}||}\eqno(25)
$$ 
for all $R > R'$ and similarly for the other
$$\eqalign{&
||(\sqrt{\mu}F_{a+t})\chi_{(R+2,\infty]}|| = \bigg[\int_{R+2}^\infty |f(x-a-t)|^2\mu(x)dx \bigg]^{1/2}=\bigg[\int_{R+2-a-t}^\infty |f(x)|^2\mu(x+a+t)dx\bigg]^{1/2}\cr
&\leq 
\bigg[\int_{R+1-a}^\infty |f(x)|^2\mu(x+n+2)dx\bigg]^{1/2} =  \bigg[\int_{R+1-a}^\infty |f(x)|^2\bigg\{ \prod_{k=0}^{n+1}{\mu(x+k+1)\over \mu(x+k)}\bigg\}\mu(x)dx\bigg]^{1/2}\cr
& \leq 
(1 + C_\mu)^{1+n/2}\bigg[\int_{R+1-a}^\infty |f(x)|^2\mu(x)dx\bigg]^{1/2}\leq (1 + C_\mu)^{1+n/2}||(\sqrt{\mu}f)\chi_{(R+1-a,\infty]}||
}\eqno(26) 
$$
for all $R > R'$. Thus we have that 
$$\eqalign{
||\sqrt{\mu}F_a - \sqrt{\mu}F_{a+t}||& \leq 
\bigg[\int_{a/2}^{R+2}|F_a(x)-F_a(x-t)|^2\mu(x)dx\bigg]^{1/2} +  
2(1 + C_\mu)^{1+n/2}||(\sqrt{\mu}f)\chi_{(R+1-a,\infty]}||}\eqno(27) 
$$
for all $R > R'$ and in consequence of $||\sqrt{\mu}f~|| < \infty$ we can always find an $R''(\epsilon)$ such that
$$
(1 + C_\mu)^{1+n/2}||(\sqrt{\mu}f)\chi_{[R,\infty]}|| < \epsilon/4\hbox{ for all }R > \max \{R''(\epsilon)+a-1, R'\}.\eqno(28)
$$
To estimate the norm expression $||\sqrt{\mu}(F_a - F_{a+t})\chi_{[a/2,R+2]}||$ in the intial  norm inequality (24) of this proof, 
we convert this to an integral expression and observe that $F_a$ is continuous on $[0,\infty)$.  As a result, $F_a$ is uniformly continuous on the compact interval $[0,R+3]$. This guarantees that to every $\epsilon > 0$ there corresponds a $\Delta_a(\epsilon)$ such that 
$$
x,y \in [0,R+3]\hbox{ with }|x-y| < \Delta_a(\epsilon)\Rightarrow|F_a(x)- F_a(y)| < \epsilon/2.\eqno(29)
$$ 
By defining the non-negative number $\mu_{a,R} \equiv \max\{\mu(x):a/2 \leq x \leq R+3 \}$, we shall have under the condition $|t| < \min\{\Delta_a(\epsilon^2[4(R+2)(\mu_{a,R}+1)]^{-1}),a/4 \}$ that
$$
\bigg[\int_{a/2}^{R+2}|F_a(x)-F_a(x-t)|^2\mu(x)dx\bigg]^{1/2} < \epsilon/2.\eqno(30)
$$
Choosing first in inequality (28) an $R > \max \{R''(\epsilon)+a-1, R'\}$ and thereafter calculating the corresponding $\mu_{a,R}$, we have for the sum of the norms
$$
||\sqrt{\mu}(F_a - F_{a+t})\chi_{[a/2,R+2]}|| + ||(\sqrt{\mu}F_a)\chi_{(R+2,\infty)}|| + ||(\sqrt{\mu}F_{a+t})\chi_{(R+2,\infty)}|| < \epsilon,\eqno(31)
$$ 
whereby the proof is complete.\qed
\medskip
\noindent{\bf LEMMA 4.4:}
For $f \in D(T_{2,F})$ satisfying $\sqrt{\mu}f \in L_2(0,\infty)$, we have for each $a >0$ that $||\sqrt{\mu}(F_a - F_a\ast\delta_\eta)||\rightarrow 0$ as $\eta \rightarrow 0^+$.
\medskip
\noindent{PROOF.}
We shall only admit $\eta < a/8$ and return to inequality (4), apply to it Jensen's integral inequality
(convexity of the function $u^2$) which yields for us
$$
|F_a(x) - (F_a\ast \delta_\eta)(x)|^2\mu(x) \leq \int_{-\eta}^\eta |F_a(x) - F_a(x - t)|^2\mu(x)\delta_\eta(t)dt,\eqno(32)
$$
and thus leads us, by means of the Minkowski integral inequality, to
$$
\bigg[\int_0^\infty |F_a(x) - (F_a\ast \delta_\eta)(x)|^2\mu(x)dx\bigg]^{1/2} \leq \int_{-\eta}^\eta \bigg[\int_0^\infty|F_a(x) - F_a(x - t)|^2\mu(x)dx\bigg]^{1/2}\delta_\eta(t)dt.\eqno(33)
$$
We convert both of the expressions in inequality (33), by bringing $\sqrt{\mu(x)}$ underneath the absolute values raised to the power $2$, to the simple norm expression
$$
||\sqrt{\mu}(F_a - F_a\ast\delta_\eta)|| \leq \int_{-\eta}^\eta ||\sqrt{\mu}(F_a - F_{a+t})||\delta_\eta(t)dt.\eqno(34)
$$
Because by Lemma 4.3, $||\sqrt{\mu}(F_a -F_{a+t})||\rightarrow 0$ for $t\rightarrow 0$ for $a > 0$, we are guaranteed the existence of a $\delta''(\epsilon)$ such that from $|t| < \delta''(\epsilon)$ it follows $||F_a -F_{a+t}|| < \epsilon$. This entails, by the immediately preceeding inequality (34), that $||\sqrt{\mu}(F_a - F_a\ast\delta_\eta)|| < \epsilon$ for $\eta< \delta''(\epsilon)$; thereby completing the proof.\qed
\medskip
\noindent{\bf LEMMA 4.5:}
For $f \in D(T_{2,F})$ with $\sqrt{\mu}f \in L_2(0,\infty)$, $a >0$ and $0 < \eta < a/8$, we have that $||\sqrt{\mu}(F_a\ast\delta_\eta -(F_a\ast\delta_\eta)\varphi_{R,r})||\rightarrow 0$ as $R \rightarrow \infty$.
\medskip
\noindent{PROOF.}
We need only to observe that in
$$
||\sqrt{\mu}(F_a\ast\delta_\eta -(F_a\ast\delta_\eta)\varphi_{R,r})|| = ||\sqrt{\mu}(F_a\ast\delta_\eta)[1 - 
\varphi_{R,r}]||\leq ||\sqrt{\mu}(F_a\ast\delta_\eta)\chi_{[R,\infty)}||\eqno(35)
$$
where the last norm expression tends towards zero as $R\rightarrow \infty$.\qed
\medskip
We now have all the tools necessary for ascertaining the adjoint ${H_\lambda}^\dagger$ of the operator $H_\lambda$ in $L_2(0,\infty)$ in terms of $D({H_\lambda}^\dagger)$, as well as its action ${H_\lambda}^\dagger:D({H_\lambda}^\dagger)\rightarrow L_2(0,\infty)$. 
\medskip
\noindent {\bf 5. The adjoint of $H_\lambda$}\medskip
We already have the adjoint ${{\bf H}_\lambda}^\dagger$ of ${\bf H}_\lambda$ in the 
Hilbert space $L_2(0,\infty)$, because the validity of the inner product equation 
$< {\bf H}_\lambda f\ |\ g > = < f\ |\ {{\bf H}_\lambda}^\dagger g >$ was easy to 
establish for all $f \in D({\bf H}_\lambda)$ and $g \in D({{\bf H}_\lambda}^\dagger)$. 
This was so, because $D({\bf H}_\lambda) = C_0^\infty(0,\infty)\cap D(M_{\lambda,\alpha})$ and $C_0^\infty(0,\infty)$ 
permitted a direct way of dealing with the integration by parts formula 
(supports of $C_0^\infty(0,\infty)$-functions are compact subsets of $(0,\infty)$) in spite of the fact that $g \in D({{\bf H}_\lambda}^\dagger)$ guarantees merely that $g'' \in L_2^{loc}(0,\infty)$ and $g' \in A(0,\infty)$ without any indication as to whether $g'(0^+)$ even exists.

In consequence of the previous section dealing with special density properties of $C_0^\infty(0,\infty)$, in particular the three triplet statements (4.2), (4.7) and (4.14), the four lemmas, Lemma 4.2, 4.3, 4.4, 4.5, provide the desired capability of approximating $D(H_\lambda)$-functions by means $C_0^\infty(0,\infty)$-functions. Hence, we shall proceed to determine the adjoint ${H_\lambda}^\dagger$ of $H_\lambda$, which process depends upon 
\medskip
\noindent{\bf THEOREM 5.1:} To every $f \in D(H_\lambda) = D(T_{2,F})\cap D(M_{\lambda,\alpha})$ and $\epsilon > 0$ 
there  exists an $f_\epsilon \in C_0^\infty(0,\infty)$ such that
$$
||M_{\lambda,\alpha}[f - f_\epsilon]||,\ ||f^{(j)} - {f_\epsilon}^{(j)}|| < \epsilon\ (j=0,1,2).\eqno(1)
$$
\noindent{PROOF.}
We first note that the positive continuous function $M_{\lambda,\alpha}(x) = [bx^2 + Ax^{-2} + \lambda x^{-\alpha}]$, 
defining the maximal multiplication operator $M_{\lambda,\alpha}$ in the Hilbert space $L_2(0,\infty)$, is strictly 
decreasing and strictly increasing on the intervals $(0,x_0]$ and $[x_0,\infty)$ respectively, where 
$x_0 = x_0(A,B,\alpha;\lambda)$. Futhermore, $M_{\lambda,\alpha}(x)/M_{\lambda,\alpha}(x+1)\rightarrow 1$ as 
$x\rightarrow \infty$ from elementary limit calculation. Therefore we conclude that 
$\mu(x) \equiv {M_{\lambda,\alpha}}^2(x) = [bx^2 + Ax^{-2} + \lambda x^{-\alpha}]^2$ is an admissible 
function for each of the  Lemmas 4.2, 4.4 and 4.5 of section 4. 
Therefore, we are justified in modifying the three triplet function statemets ((4.2), (4.7) and (4.14)) by adding in the limit 
statements of Lemmas 4.2, 4.4 and 4.5 to the limit statements of the first, second and third triplet for $\mu(x) \equiv {M_{\lambda,\alpha}}^2(x)$ respectively. Thus given an arbitrary $\epsilon > 0$, we consider first an $a >0$ such that for the first triplet 
$\{F_a, {F_a}', {F_a}'' \}$ of functions
$$ 
||M_{\lambda,\alpha}[f - F_a]||,\ ||f^{(j)} - {F_a}^{(j)}|| < \epsilon/3\ (j=0,1,2)\eqno(2) 
$$
holds. Thereafter we choose an $\eta < a/8$ so that for the second triplet  $\{(F_a\ast\delta_\eta) , (F_a\ast\delta_\eta)', (F_a\ast\delta_\eta)'' \}$
of functions 
$$
||M_{\lambda,\alpha}[F_a - F_a\ast\delta_\eta]||,\ ||{F_a}^{(j)} - ({F_a}\ast \delta_\eta)^{(j)}|| < \epsilon/3\ (j=0,1,2).\eqno(3) 
$$ 
becomes valid. And we finally pick an $R$ so large that for the third triplet $\{(F_a\ast\delta_\eta)\varphi_{R,r} , ((F_a\ast\delta_\eta)\varphi_{R,r})', ((F_a\ast\delta_\eta)\varphi_{R,r})'' \}$ of functions  
$$
||M_{\lambda,\alpha}[F_a\ast\delta_\eta - (F_a\ast\delta_\eta)\varphi_{R,r}]||,\ ||(F_a \ast \delta_\eta)^{(j)} - ((F_a \ast \delta_\eta)\varphi_{R,r})^{(j)}|| < \epsilon/3\ (j=0,1,2)\eqno(4)
$$
is guaranteed. By writing $f_\epsilon \equiv (F_a\ast\delta_\eta)\varphi_{R,r}$ we obtain a $C_0^\infty(0,\infty)$-function for which 
$$
||M_{\lambda,\alpha}[f - f_\epsilon]||,\ ||f^{(j)} - {f_\epsilon}^{(j)}|| < \epsilon\ (j=0,1,2)\eqno(5)
$$
and thereby completing our proof.\qed
\medskip
\noindent{\bf THEOREM 5.2:} For all $f \in D(H_\lambda) = D(T_{2,F})\cap D(M_{\lambda,\alpha})$ and $g \in D({{\bf H}_\lambda}^\dagger)$ we have that 
$$
< H_\lambda f\ |\ g > = < f\ |\ {{\bf H}_\lambda}^\dagger g >.\eqno(6)
$$
\medskip
\noindent{PROOF.}
We apply Theorem 5.1 for an arbitrary $\epsilon > 0$ - i. e. there exists an $f_\epsilon \in C_0^\infty(0,\infty)$ such that 
$||M_{\lambda,\alpha}[f - f_\epsilon]|| < \epsilon$ and $||f'' - {f_\epsilon}''|| < \epsilon$. We calculate and re-arrange
$$\eqalign{&
\bra\ H_\lambda f\ |\ g\ \ket = \bra\ -f'' + M_{\lambda,\alpha}f\ |\ g\ \ket
= \bra\ -f''\ |\ g\ \ket + \bra\ M_{\lambda,\alpha}f\ |\ g\ \ket \cr
& = 
\bra\ -[f'' - {f_\epsilon}''] - {f_\epsilon}''\ |\ g\ \ket + \bra\ M_{\lambda,\alpha}[f - f_\epsilon] + M_{\lambda,\alpha}f_\epsilon\ |\ g\ \ket \cr
&= 
- \bra\ f'' - {f_\epsilon}''\ |\ g\ \ket + \bra\ M_{\lambda,\alpha}[f - f_\epsilon]\ |\ g\ \ket + \bra\ {\bf H}_\lambda f_\epsilon\ |\ g\ \ket \cr
&=
- \bra\ f'' - {f_\epsilon}''\ |\ g\ \ket + \bra\ M_{\lambda,\alpha}[f - f_\epsilon]\ |\ g\ \ket + \bra\ f_\epsilon\ |\ {{\bf H}_\lambda}^\dagger g\ \ket \cr
&=
- \bra\ f'' - {f_\epsilon}''\ |\ g\ \ket + \bra\ M_{\lambda,\alpha}[f - f_\epsilon]\ |\ g\ \ket - \bra\ f - f_\epsilon\ |\ {{\bf H}_\lambda}^\dagger g\ \ket + \bra\ f\ |\ {{\bf H}_\lambda}^\dagger g\ \ket,}\eqno(7)                                         
$$
which chain of equalities implies via the Cauchy-Schwarz inequality that
$$\eqalign{&
|\bra\ H_\lambda f\ |\ g\ \ket - \bra\ f\ |\ {{\bf H}_\lambda}^\dagger g\ \ket|  \leq
|- \bra\ f'' - {f_\epsilon}''\ |\ g\ \ket + \bra\ M_{\lambda,\alpha}[f - f_\epsilon]\ |\ g\ \ket \cr & - \bra\ f - f_\epsilon\ |\ {{\bf H}_\lambda}^\dagger g\ \ket| \leq
|\bra\ f'' - {f_\epsilon}''\ |\ g\ \ket| + |\bra M_{\lambda,\alpha}[f - f_\epsilon]\ |\ g\ \ket| + |\bra\ f - f_\epsilon\ |\ {{\bf H}_\lambda}^\dagger g\ \ket|\cr
& \leq
\|f'' - {f_\epsilon}''\|\times\|g\| + \|M_{\lambda,\alpha}[f - f_\epsilon]\|\times\|g\| + \|f - f_\epsilon\|\times\|{{\bf H}_\lambda}^\dagger g\| < \epsilon [2\|g\| + \|{{\bf H}_\lambda}^\dagger g\|].}\eqno(8)
$$
However, $\epsilon$ was an arbitrary positive number, therefore we conclude that
$$
< H_\lambda f\ |\ g > = < f\ |\ {{\bf H}_\lambda}^\dagger g >\hbox{ for all }f \in D(H_\lambda)\hbox{ and }g \in D({{\bf H}_\lambda}^\dagger),\eqno(9)
$$
thereby ending the proof.\qed
\medskip
\noindent{\bf THEOREM 5.3:}
For the perturbed Hamiltonian $H_\lambda$, we have that ${H_\lambda}^\dagger = {{\bf H}_\lambda}^\dagger$.
\medskip
\noindent{PROOF.}
The inner product statement of Theorem 5.2 implies that $H_\lambda$ and ${{\bf H}_\lambda}^\dagger$ are formally adjoint to each other, and thus ${H_\lambda}^\dagger \supset {{\bf H}_\lambda}^\dagger$. On the other hand, out of $H_\lambda \supset {\bf H}_\lambda$ shall follow ${{\bf H}_\lambda}^\dagger\supset{H_\lambda}^\dagger$, and our proof is thus completed.\qed
\medskip
Having obtained the adjoint ${H_\lambda}^\dagger$ of the perturbed Hamiltonian $H_\lambda$, we may now turn our attention to the construction of the Friedrichs extension $T_\lambda$ of $H_\lambda$ in the Hilbert space $L_2(0,\infty)$. 

\medskip
\noindent {\bf 6. The Friedrichs Extension of $H_\lambda$}
\medskip
Prior to embarking upon the derivation of the Friedrichs extensions of the $\lambda$-parameter family $H_\lambda$ of operators, 
we must return to the triplets at the beginning of Section 4 ((4.2), (4.7) and (4.14)) and modify these triplets to pairs by dropping 
the second derivatives. We recall that for every $f \in D(T_{2,F})$, we obtained for each of the three triplets 
$\{F_a, {F_a}', {F_a}'' \}$, $\{(F_a\ast \delta_\eta), (F_a\ast \delta_\eta)', (F_a\ast \delta_\eta)''\}$ and 
$\{(F_a\ast \delta_\eta)\varphi_{R,r}, ((F_a\ast \delta_\eta)\varphi_{R,r})',  ((F_a\ast\delta_\eta)\varphi_{R,r})''\}$ 
the following limit statments $(j=0,1,2)$:
$$||f^{(j)}- {F_a}^{(j)}|| \rightarrow 0,\ ||{F_a}^{(j)} - (F_a\ast \delta_\eta)^{(j)}|| \rightarrow 0~
,\ ||(F_a\ast \delta_\eta)^{(j)} - ((F_a\ast \delta_\eta)\varphi_{R,r})^{(j)}|| \rightarrow 0 $$
for $a\rightarrow 0^+, \eta \rightarrow 0^+,$ and $R \rightarrow \infty$ respectively.
These limit statements can be simply modified for absolutely continuous $L_2(0,\infty)$-functions $f$ with $f(0)=0$ and 
$f' \in L_2(0,\infty)$. If $f$ is such a function, then $F_a$, defined by $F_a(x) \equiv 0$ for $0 \leq x < a$ and $\equiv f(x-a)$ 
for $x \geq a$, is an absolutely continuous $L_2(0,\infty)$-function with $F_a(0)=0$ and ${F_a}' \in L_2(0,\infty)$ for all 
$a > 0$. All of these $||\cdot||$-limit statements came about by means of: the Jensen integral inequality for the measure $\delta_\eta(x)dx$
 on $(-\eta,\eta)$ combined with the Tonnelli-Hobson Theorem, the Minkowski integral inequality, as well as the integration by parts 
formula going up to the second derivative. Therefore, we treat absolutely continuous $L_2(0,\infty)$-function $f$ with $f(0)=0$ and 
$f' \in L_2(0,\infty)$ likewise, however we go up to the first derivative only. Hence, we obtain analoguous $||\cdot||$-limit statements 
for the three pairs $\{F_a, {F_a}'\}, \{(F_a\ast \delta_\eta), (F_a\ast \delta_\eta)'\}$ and $\{(F_a\ast \delta_\eta)\varphi_{R,r}, 
((F_a\ast \delta_\eta)\varphi_{R,r})'\}$ as before, and these are as follows $(j=0,1)$:
$$||f^{(j)}- {F_a}^{(j)}|| \rightarrow 0,
	||{F_a}^{(j)} - (F_a\ast \delta_\eta)^{(j)}|| \rightarrow 0,
	||(F_a\ast \delta_\eta)^{(j)} - ((F_a\ast \delta_\eta)\varphi_{R,r})^{(j)}|| \rightarrow 0 $$
for $a\rightarrow 0^+$, $\eta \rightarrow 0^+$ and  $R \rightarrow \infty$ respectively. 
At this stage we point to the fact that the function $M_{\lambda,\alpha}(x) = [Bx^2 + Ax^{-2} + \lambda x^{-\alpha}]$, 
which is continuous on $(0,\infty)$, satisfies the conditions required by the function $\mu(x)$ appearing in Lemma 4.5. 
Therefore, if we assume for our absolutely continuous $L_2(0,\infty)$-function $f$, with $f(0)=0$ and $f' \in L_2(0,\infty)$, 
the further property of $\sqrt{M_{\lambda,\alpha}}f \in L_2(0,\infty)$, then we shall have in addition to the immediately preceeding 
$||\cdot||$-limit statements for the given pairs the following:
$$||\sqrt{M_{\lambda,\alpha}}[f - F_a]|| \rightarrow 0, ||\sqrt{M_{\lambda,\alpha}}[F_a - F_a\ast \delta_\eta]|| \rightarrow 0,
||\sqrt{M_{\lambda,\alpha}}[F_a\ast \delta_\eta - (F_a\ast \delta_\eta)\varphi_{R,r}]|| \rightarrow 0,
$$for $a\rightarrow 0^+$, $\eta \rightarrow 0^+$ and  $R \rightarrow \infty$ respectively. We have consequently arrived at the following 
\medskip
\noindent{\bf LEMMA 6.1:}
If $f$ is an absolutely continuous $L_2(0,\infty)$-function $f$ with $f(0) = 0$ and $f'$ and $\sqrt{M_{\lambda,\alpha}}f \in L_2(0,\infty)$, then for $j=0$ and $1$ we have that 
$$
||\sqrt{M_{\lambda,\alpha}}[F_a - F_a\ast \delta_\eta]||, ||f^{(j)}- {F_a}^{(j)}|| \rightarrow 0\hbox{ for }a\rightarrow 0^+;\eqno(1) 
$$
$$
||\sqrt{M_{\lambda,\alpha}}[F_a - F_a\ast \delta_\eta]||, ||{F_a}^{(j)} - (F_a\ast \delta_\eta)^{(j)}|| \rightarrow 0\hbox{ for }\eta \rightarrow 0^+;\eqno(2)
$$
$$
||\sqrt{M_{\lambda,\alpha}}[F_a\ast \delta_\eta - (F_a\ast \delta_\eta)\varphi_{R,r}]||,
||(F_a\ast \delta_\eta)^{(j)} - ((F_a\ast \delta_\eta)\varphi_{R,r})^{(j)}|| \rightarrow 0\hbox{ for }R \rightarrow \infty.\eqno(3)
$$

\noindent These $||\cdot||$-limit statements permit us to fomulate a theorem analoguous to Theorem 4, namely
\medskip
\noindent{\bf THEOREM 6.2:} To every $f \in A(0,\infty)\cap L_2(0,\infty)$ with $f(0) = 0$, $f'$ and $\sqrt{M_{\lambda,\alpha}}f \in L _2(0,\infty)$ and pre-assigned $\epsilon > 0$, there exists an $f_\epsilon \in C_0^\infty(0,\infty)$ such that
$$
||\sqrt{M_{\lambda, \alpha}}[f - f_\epsilon]||, ||f - {f_\epsilon}^{(j)}|| < \epsilon\ (j=0,1).\eqno(4) 
$$
\noindent{PROOF.}
We shall apply Lemma 6.1 by first picking an $a > 0$ satisfying $||\sqrt{M_{\lambda,\alpha}}[F_a - F_a\ast \delta_\eta]||, ||f^{(j)}- {F_a}^{(j)}|| < \epsilon/3$ $(j=0,1)$. Thus we find an $\eta > 0$ so small that $\eta < a/8$, for which $||\sqrt{M_{\lambda,\alpha}}[F_a - F_a\ast \delta_\eta]||, ||{F_a}^{(j)} - (F_a\ast \delta_\eta)^{(j)}|| < \epsilon/3$ $(j=0,1)$. And finally, we choose an $R$ so large that $||\sqrt{M_{\lambda,\alpha}}[F_a\ast \delta_\eta - ((F_a\ast \delta_\eta)\varphi_{R,r})]||, ||(F_a\ast \delta_\eta)^{(j)} - ((F_a\ast \delta_\eta)\varphi_{R,r})^{(j)}|| < \epsilon/3$ $(j=0,1)$. Herewith we define $f_\epsilon \equiv (F_a\ast \delta_\eta)\varphi_{R,r} \in C_0^\infty(0,\infty)$, which, by the choice $\epsilon/3,$ gives the concluding statements of this theorem, thus ending the proof.\qed
\medskip
\noindent We want to emphasize that in the above theorem, the approximating $C_0^\infty(0,\infty)$-function $f_\epsilon$ also belongs to $D(H_\lambda) = D(T_{2,F})\cap D(M_{\lambda,\alpha})$. 

To arrive at the Friedrichs extensions $T_\lambda$ of the $\lambda$-parameter family $H_\lambda$ of operators, we have to return individually to each of the semi-bounded operators $H_\lambda$ in the Hilbert space $L_2(0,\infty)$, which is each  bounded below by $\gamma = \gamma(A,B,\lambda;\alpha) \equiv \min M_{\lambda, \alpha}((0,\infty))$. They generated a $\lambda$-parameter family of semi-bounded sesquilinear forms $s_\lambda$ on $D(H_\lambda)\times D(H_\lambda)$, having lower bound $\gamma = \gamma(A,B,\lambda;\alpha)$, defined by
$$
s_\lambda(f,g) \equiv < H_\lambda f\ |\ g > = < -f''\ |\ g > + < M_{\lambda, \alpha}f\ |\ g >\hbox{ for all }f,g \in D(H_\lambda).\eqno(5)
$$
Herein we may write $< -f''\ |\ g > = < f'\ |\ g' >$ in consequence of 
$$\eqalign{
< -f''\ |\ g > &= \lim_{R\rightarrow \infty}\int_0^R -f''(x)\overline{g(x)}dx = \lim_{R\rightarrow \infty}\bigg[f'(x)\overline{g'(x)}|_0^R + \int_0^R f'(x)\overline{g'(x)}dx\bigg] \cr
&=
\lim_{R\rightarrow \infty}\bigg[f'(R)\overline{g'(R)} + \int_0^R f'(x)\overline{g'(x)}dx\bigg]= 
\int_0^\infty f'(x)\overline{g'(x)}dx= < f'\ |\ g' >,}
$$
where we have used the fact that $f,g \in D(H_\lambda) = D(T_{2,F})\cap D(M_{\lambda, \alpha})$ implies $f^{(j)}(0)=g^{(j)}(0)=0$ as well as $f^{(j)}(R),g^{(j)}(R)\rightarrow 0$ as $R\rightarrow \infty\ (j=0,1)$ since \sref{\jowa, p 153, Theorem 6.27} $f,g \in W_{2,2}(0,\infty)$. We may therefore replace $< -f''\ |\ g >$ with $< f'\ |\ g' >$ in the definition of $s_\lambda$, and thus have that 
$$
s_\lambda(f,g) \equiv < H_\lambda f\ |\ g > = < f'\ |\ g' > + <\sqrt{M_{\lambda, \alpha}}f | \sqrt{M_{\lambda, \alpha}}g >\hbox{ for all }f,g \in D(H_\lambda)\eqno(6)
$$
is a semi-bounded sesquilinear form bounded below by $\gamma = \gamma(A,B,\lambda;\alpha)$, which induces the $\lambda$-parameter family of inner products 
$$\eqalign{
< f\ |\ g >_\lambda &\equiv s_\lambda(f,g) + [1-\gamma]< f\ |\ g > \cr
&= 
< f'\ |\ g' > + <\sqrt{M_{\lambda, \alpha}}f\ |\ \sqrt{M_{\lambda, \alpha}}g > + [1-\gamma]
< f\ |\ g >;}\eqno(7)
$$
on $ D(H_\lambda)$, in other words, $(D(H_\lambda),< \cdot\ |\ \cdot >_\lambda)$ is a Pre-Hilbert space, whose norm 
$$
||f||_\lambda \equiv \sqrt{||f'||^2 + ||\sqrt{M_{\lambda, \alpha}}f||^2 + [1-\gamma]||f||^2}\  \ \bigg( \geq  ||f|| \bigg)\eqno(8)
$$
is compatible with the original $L_2(0,\infty)$-norm $||\cdot||$. This follows from the fact that each of the sesquilinear forms $s_\lambda$ was defined \sref{\jowa, p 119}-\sref{\rina, p 329, Sec. 124} by the semi-bounded operator $H_\lambda$ in the Hilbert space  $L_2(0,\infty)$. Because of this, we may consider the completion $({\cal H}_\lambda,< \cdot\ |\ \cdot >_\lambda)$ of $(D(H_\lambda),< \cdot\ |\ \cdot >_\lambda)$ as a subspace of $L_2(0,\infty)$. As a result of this, every $f\in{\cal H}_\lambda$ must be absolutely continuous with $||f'||^2 + ||\sqrt{M_{\lambda, \alpha}}f||^2 + [1-\gamma]||f||^2 < \infty $ and moreover $f(0)=0$, because any sequence $\{f_n\}$ from $D(H_\lambda)$ converging to $f \in{\cal H}_\lambda$ in norm $||\cdot||_\lambda$ satisfies 
$$
f_n(x) = \int_0^x {f_n}'(t)dt \rightarrow \int_0^x f'(t)dt = f(x)\hbox{ as }n\rightarrow \infty.
$$
This lets us thoroughly describe the completion ${\cal H}_\lambda$ of $D(H_\lambda)$ in terms of the norm $||\cdot||_\lambda$ as
$$
{\cal H}_\lambda = \{f\in L_2(0,\infty): ||f'||^2 + ||\sqrt{M_{\lambda, \alpha}}f||^2 + [1-\gamma]||f||^2 < \infty\}  
$$
and these subspaces of $L_2(0,\infty)$ we formulate more elegantly in terms of 
\medskip
\noindent{\bf THEOREM 6.3:} The completions ${\cal H}_\lambda$ of $D(H_\lambda)$ in terms of the respective norms $||\cdot||_\lambda$ is
$$
\{f\in L_2(0,\infty):f \in A(0,\infty), f(0)=0,\hbox{ both }f'\hbox{ and }\sqrt{M_{\lambda, \alpha}}f \in L_2(0,\infty)\}.\eqno(9)   
$$
\medskip
\noindent{PROOF.}
The completion ${\cal H}_\lambda$ is clear from the above. Therefore, let us assume that $f \in L_2(0,\infty)$ satisfies all the conditions specified within the curly brackets. As consequence of Theorem 6.2 we can find a sequence of $C_0^\infty(0,\infty)$-functions $\{f_n\}$ such that $||\sqrt{M_{\lambda, \alpha}}[f-f_n]||, ||f-f_n||$ and $||f'-{f_n}'||\rightarrow 0$ as $n\rightarrow \infty$, which sequence also belongs to $D(H_\lambda) = D(T_{2,F})\cap D(M_{\lambda,\alpha})$. This sequence is a $||\cdot||_\lambda$-Cauchy sequence from $D(H_\lambda)$ with the property that
$$
||f-f_n||_\lambda = \sqrt{||f'-{f_n}'||^2 + ||\sqrt{M_{\lambda, \alpha}}[f-f_n]||^2 + [1-\gamma]||f-f_n||^2} \rightarrow 0\hbox{ as }n\rightarrow \infty\hbox{ and }
$$     
$$
f_n(x) = \int_0^x {f_n}'(t)dt \rightarrow \int_0^x f'(t)dt = f(x)\hbox{ as }n\rightarrow \infty\hbox{ for all }x \geq 0.
$$ 
Thereby the proof is complete.\qed

Finally, we come to the construction of the Friedrichs extension $T_\lambda$ of perturbed Hamiltonian operator $H_\lambda$ in the Hilbert space $L_2(0,\infty)$. As is well established \sref{\jowa, p 120, Theorem 5.38}-\sref{\rina, p 335}, the operator $T_\lambda$ has domain of definition $D(T_\lambda) = D({H_\lambda}^\dagger)\cap {\cal H}_\lambda$ and action $T_\lambda f \equiv {H_\lambda}^\dagger f$ for all $f \in D(T_\lambda)$. As a direct result of the immediately preceeding paragraph and Theorem 6.3, we may now formulate
\medskip
\noindent{\bf THEOREM 6.4}
The Friedrichs extension $T_\lambda$ of the perturbed Hamiltonian operator $H_\lambda$ has domain of definition and action given by
$$
D(T_\lambda) = \{f \in L_2(0,\infty):(-f'' + M_{\lambda,\alpha}f) \in L_2(0,\infty), f(0)=0, f \in A(0,\infty)\hbox{ with }
$$
$$
\hbox{ both }f'\hbox{ and }\sqrt{M_{\lambda,\alpha}}f \in L_2(0,\infty)\}\hbox{ and }T_\lambda f = {H_\lambda}^\dagger f\hbox{ for all }f \in D(T_\lambda).\eqno(10)
$$
\medskip
\noindent {\bf Acknowledgments}
\medskip
\noindent Partial financial support of this work under Grant Nos. GP249507 and GP3438 from the 
Natural Sciences and Engineering Research Council of Canada is gratefully 
acknowledged by two of us [NS, RLH].\medskip

\references{1}

\end